\documentclass[aps,prd,showkeys,showpacs,twocolumn,superscriptaddress]{revtex4}
\usepackage[dvips]{graphicx}
\usepackage{amsfonts} 
\usepackage{amsmath} 
\usepackage{latexsym} 
\usepackage{amssymb}

\usepackage{color}

\usepackage{eurosym}

\newcommand{\ket}[1]{| #1 \rangle}
\newcommand{\bra}[1]{\langle #1 |}

\newcommand{\trace}{\textrm{Tr}}

\newcommand{\couic}[1]{}

\newcommand{\Id}{\operatorname{Id}}
\newcommand{\C}{\mathbb{C}}
\newcommand{\Z}{\mathbb{Z}}
\newcommand{\joliA}{\mathcal{A}}
\newcommand{\joliB}{\mathcal{B}}
\newcommand{\joliC}{\mathcal{C}}
\newcommand{\joliH}{\mathcal{H}}
\newcommand{\joliL}{\mathcal{L}}

\newcommand{\joliN}{\mathcal{N}}
\newcommand{\pa}[1]{\left(#1\right)}

\newtheorem{Def}{Definition}
\newtheorem{Th}{Theorem}

\newtheorem{Lem}{Lemma}

\newtheorem{Pro}{Proposition}

\begin{document}

\title{\begin{center}
One-dimensional quantum cellular automata over finite, unbounded configurations.
\end{center}}

\author{Pablo Arrighi}
\email{pablo.arrighi@imag.fr} 
\affiliation{Universit\'e de Grenoble,\\ 
LIG, 46 Avenue F\'elix Viallet, 38031 Grenoble Cedex, France. }

\author{Vincent Nesme}
\email{vincent.nesme@lri.fr}

\author{Reinhard Werner}
\email{R.Werner@tu-bs.de} 
\affiliation{Technical University of Braunschweig,\\ 
IMAPH, Mendelssohnstr. 3, 38106 Braunschweig, Germany. }

\begin{abstract}
One-dimensional quantum cellular automata (QCA) consist in a line of identical, finite dimensional quantum systems. These evolve in discrete time steps according to a local, shift-invariant unitary evolution. By local we mean that no instantaneous long-range communication can occur. In order to define these over a Hilbert space we must restrict to a base of finite, yet unbounded configurations. We show that QCA always admit a two-layered block representation, and hence the inverse QCA is again a QCA. This is a striking result since the property does not hold for classical one-dimensional cellular automata as defined over such finite configurations. As an example we discuss a bijective cellular automata which becomes non-local as a QCA, in a rare case of reversible computation which does not admit a straightforward quantization. We argue that a whole class of bijective cellular automata should no longer be considered to be reversible in a physical sense. Note that the same two-layered block representation result applies also over infinite configurations, as was previously shown for one-dimensional systems in the more elaborate formalism of operators algebras \cite{Werner}. Here the proof is made simpler and self-contained, moreover we discuss a counterexample QCA in higher dimensions.
\end{abstract}

\keywords{Symbolic dynamics, complex systems, intrinsic universality, complex systems, translation-invariant dynamics, Ising models}

\pacs{03.67.Lx, 03.67.-a, 03.65.-w, 05.30.-d}

\maketitle

One-dimensional cellular automata (CA) consist in a line of cells, each of which may take one in a finite number of possible states. These evolve in discrete time steps according to a local, shift-invariant function. When defined over infinite configurations, the inverse of a bijective CA is then itself a CA, and this structural reversibility leads to a natural block decomposition of the CA. None of this holds over finite, yet possibly unbounded, configurations. 

Because CA are a physics-like model of computation it seems very natural to study their quantum extensions. The flourishing research in quantum information and quantum computer science provides us with appropriate context for doing so, both in terms of the potential implementation and the theoretical framework. Right from the very birth of the field with Feynman's 1986 paper, it was hoped that QCA may prove an important path to realistic implementations of quantum computers \cite{Feynman} -- mainly because they eliminate the need for an external, classical control and hence the principal source of decoherence. Other possible aims include providing models of distributed quantum computation, providing bridges between computer science notions and modern theoretical physics, or anything like understanding the dynamics of some quantum physical system in discrete spacetime, i.e. from an idealized viewpoint. Studying QCA rather than quantum Turing machines for instance means we bother about the spatial structure of things \cite{Arrighi2}, whether for the purpose of describing a quantum protocol, modelling a quantum physical phenomena, or again taking into account the spatial parallelism inherent to the model. 

One-dimensional quantum cellular automata (QCA) consist in a line of identical, finite dimensional quantum systems. These evolve in discrete time steps according to a local, shift-invariant unitary evolution. By local we mean that information cannot be transmitted faster that a fixed number of cells per time step. Because the standard mathematical setting for quantum mechanics is the theory of Hilbert spaces, we must exhibit and work with a countable basis for our vectorial space. This is the reason why we restrict to finite, unbounded configurations. An elegant alternative to this restriction is to abandon Hilbert spaces altogether and use the more abstract mathematical setting of $C^*$-algebras \cite{Bratteli} -- but here we seek to make our proofs self-contained and accessible to a wider community, including those computer scientists with an interest in quantum computation. Our main result is that QCA can always be expressed as two layers of an infinitely repeating unitary gate even over such finite configurations. The existence of such a two-layered block representation implies of course that the inverse QCA is again a QCA. Our result is mainly a simplification of the same theorem over infinite configurations as expressed with operators algebra \cite{Werner}. Unfortunately an example QCA disproves the theorem in further dimensions -- at least in its present form.

It is a rather striking fact however that QCA admit the two-layered block representation in spite of their being defined over finite, unbounded configurations. For most purposes this saves us from complicated unitary tests such as \cite{Durr1, Durr2, Arrighi1}. But more importantly notice how this is clearly not akin to the classical case, where a CA may be bijective over such finite configurations, and yet not structurally reversible. In order to clarify this situation we consider a perfectly valid, bijective CA but whose inverse function is not a CA. It then turns out that its quantum version is no longer valid, as it allows superluminal signalling. Hence whilst we are used to think that any reversible computation admits a trivial quantization, this turns out not to be case in the realm of cellular automata. Curiously the non-locality of quantum states (entanglement) induces more structure upon the cellular automata -- so that its evolution may remain local as an operation (no-superluminal signalling). Based upon these remarks we prove that an important, well-studied class of bijective CA may be dismissed as not physically reversible.

\emph{Outline.} We  reorganize a number of known mathematical results around the notion of subsystems in quantum theory (Section \ref{subsystems}). Thanks to this small theory we prove the reversibility/block structure theorem in an elementary manner (Section \ref{blocks}). In the discussion we show why the theorem does not hold as such in further dimensions; we exhibit superluminal signalling in the XOR quantum automata, and end with a general theorem discarding all injective, non surjective CA over infinite configurations as unphysical (Section \ref{discussion}).

\section{A small theory of subsystems}\label{subsystems}

\begin{Def}[Algebras]~\\
Consider $\mathcal{A}\subseteq M_n(\mathbb{C})$. 
We say that $\mathcal{A}$ is an \emph{algebra} of $M_n(\mathbb{C})$ if and only if
it is closed under weighting by a scalar ($.$), addition ($+$), 
matrix multiplication ($*$), adjoint ($\dagger$). 
Moreover for any $S$ a subset of $M_n(\mathbb{C})$, 
we denote by curly $\mathcal{S}$ its closure under the above-mentioned operations.
\end{Def}
Note that algebras as above-defined are really just $C^*$-algebras over finite-dimensional systems.

\begin{Def}[Subsystem algebras]~\\
Consider $\mathcal{A}$ an algebra of $M_n(\mathbb{C})$. We say that 
$\mathcal{A}$ is a \emph{subsystem algebra} of $M_n(\mathbb{C})$ if and only if 
there exists $p,q\in\mathbb{N}\,/\,pq=n$ 
and $U\in M_n(\mathbb{C})\,/\,U^{\dagger}U=UU^{\dagger}=\mathbb{I}$ 
such that $U\mathcal{A}U^{\dagger}=M_p(\mathbb{C})\otimes \mathbb{I}_q$.
\end{Def}

\begin{Def}[Center algebras]~\\
For $\joliA$ an algebra of $M_n(\C)$, we note $\joliC_\joliA=\{A\in\mathcal{A}\;|\;\forall B\in\mathcal{A}\;BA=AB\}$. $\joliC_\joliA$ is also an algebra of $M_n(\C)$, which is called the \emph{center algebra} of $\joliA$.
\end{Def}

\begin{Th}[Characterizing one subsystem]~\\  \label{subsystem}
Let $\mathcal{A}$ be an algebra of $M_n(\mathbb{C})$ and 
$\mathcal{C}_\mathcal{A}=\{A\in\mathcal{A}\;|\;\forall B\in\mathcal{A}\;BA=AB\}$
its \emph{center algebra}. Then $\mathcal{A}$ is a subsystem algebra if and only if
$\mathcal{C}_\mathcal{A}=\mathbb{C}\mathbb{I}$.
\end{Th}
\textbf{Proof.} The argument is quite technical and its understanding not mandatory for understanding the rest of the paper. Its presentation is based on \cite{Gijswijt}. See also \cite{Bratteli} for a proof within the setting of general $C^*$-algebras.\\
$[\Rightarrow]$.\\
$\mathcal{C}_{M_p(\mathbb{C})}=\mathbb{C}\mathbb{I}_p$, and hence 
$\mathcal{C}_{M_p(\mathbb{C})\otimes\mathbb{I}_q}=\mathbb{C}\mathbb{I}$.\\
$[\Leftarrow]$.\\
\textsc{Consider} some set $P=\{P_i\}_{i=1\ldots p}$ such that:\\
$(i)$  $\forall i=1\ldots p \; P_i\in\mathcal{A}$;\\
$(ii)$ $\forall i,j=1\ldots p \; P_iP_j=\delta_{ij}P_i\in\mathcal{A}$.\\
Moreover we take $P$ is maximal, i.e. so that there is no set $Q=\{Q_i\}$ 
verifying conditions $(i),(ii)$ and such that $\mathcal{P}\subset \mathcal{Q}$, with
$\mathcal{P},\mathcal{Q}$ the closures of $P,Q$.
Note that:\\
$(iii)$ $\sum_{i=1\ldots p} P_i =\mathbb{I}$,\\
otherwise $\mathbb{I}-\sum_{i=1\ldots p} P_i$ may be added to the set.\vspace{1mm}\\
\textsc{First} we show that
\begin{align}
\forall i\quad [P_i\mathcal{A}P_i=\mathbb{C}P_i].\label{diagcase}
\end{align}
Intuitively this is because the contrary would allow us to refine the subspaces defined by $P_i$ into
smaller subspaces, and hence go against the fact that $P$ is maximal. Formally consider
some $M\in\mathcal{A}$ such that $P_i M P_i\not\propto P_i$. 
If $M$ is proportional to a unitary let $N=M+M^{\dagger}$, else let $N=M^{\dagger}M$. 
In any case we have $P_i N P_i\not\propto P_i$ with $N$ hermitian.
Note that $H=P_i N P_i$ is also hermitian, 
and has support in the subspace $P_i$, hence we can write 
$H=\sum_k \lambda_k Q_k$ with the $Q_k$'s orthogonal projectors 
such that $\sum_k Q_k = P_i$ and the $\lambda_k$'s distinct real numbers. 
Any such $Q_k$ is part of $\mathcal{A}$, since 
$Q_k=\frac{H\prod_{l\neq k} (H-\lambda_l \mathbb{I})}{\lambda_k \prod_{l\neq k} \pa{\lambda_k-\lambda_i}}$.
Consider the set $Q=P/\{P_i\}\cup_k \{Q_k\}$. It satisfies condition $(i),(ii)$ but 
$\mathcal{P}\subset\mathcal{Q}$, which is impossible.\vspace{1mm}\\
\textsc{Second} we show that
\begin{align}
\forall i,j \quad [P_i\mathcal{A}P_j\neq 0].\label{nondiagcase}
\end{align}
Intuitively this is because the contrary would split $\mathcal{A}$ into the direct sum of two matrix algebras,
and hence go against the fact that $\mathcal{C}_\mathcal{A}=\mathbb{C}\mathbb{I}$. Formally write $i\sim j$ whenever this is the case. The relation $\sim$ is reflexive by Eq. (\ref{diagcase}), transitive by multiplicative closure of $\mathcal{A}$, and symmetric since 
\begin{align*} 
P_i\mathcal{A}P_j\neq 0 &\Rightarrow (P_i\mathcal{A}P_j)^{\dagger}\neq 0 \\
&\Rightarrow (P_j\mathcal{A}^{\dagger}P_i)\neq 0\\
&\Rightarrow (P_j\mathcal{A}P_i)\neq 0.
\end{align*}
Say there is an equivalence class $J\subset 1\ldots p$ and 
let $P_I =\sum_{i\notin J}P_i$, $P_J =\sum_{j\in J}P_j$. Then
\begin{align*} 
\mathcal{A}P_J &= (P_I + P_J) \mathcal{A} P_J\\
&= P_J \mathcal{A} P_J\\
&= P_J \mathcal{A}\quad\textrm{symmetrically.}
\end{align*}
and so $P_J\in\mathcal{C}_\mathcal{A}$, which is impossible.\vspace{1mm}\\
\textsc{Third} we show that for all $A$, for all $i,j=1\ldots p$, 
if we let $M=P_i A P_j$ then
\begin{align}
\exists \lambda\in\mathbb{C}\quad[M^\dagger M=\lambda P_i\,\wedge\, M M^\dagger=\lambda P_j].\label{sympropto}
\end{align}
Indeed Eq. (\ref{diagcase}) gives $M^\dagger M=\lambda P_i$ and $M M^\dagger=\mu P_j$. 
But then $\lambda^2 P_i = M^\dagger M M^\dagger M = \mu M^\dagger P_j M=  \mu\lambda P_i$, 
hence $\lambda$ equals $\mu$.\vspace{1mm}\\
\textsc{Fourth} we show that 
\begin{align}
\forall i,j \quad [\trace(P_i)=\trace(P_j)=\textrm{some constant }q].\label{qvalue}
\end{align}
For each $i,j$ take some $A\in\mathcal{A}$ verifying Eq. (\ref{nondiagcase}).
Let $M=P_i A P_j$. By Eq. (\ref{sympropto}) there is a complex number $\lambda$
such that we have $P_i=\lambda MM^{\dagger}$ and $P_j=\lambda M^{\dagger}M$.
Then the equality follows from  
$\trace(\lambda MM^{\dagger})=\trace(\lambda M^{\dagger}M)$.\vspace{1mm}\\
\textsc{Fifth} consider some unitary $U$ which takes those $\{P_i\}_{i=1\ldots p}$ 
into one-zero orthogonal diagonal matrices $I=\{\mathbf{I}_i\}_{i=1\ldots p}$ with $\mathbf{I}_i=\ket{i}\bra{i}\otimes\mathbb{I}_q$.
Note that this is always possible since the $\{P_i\}_{i=1\ldots p}$ form an 
orthogonal $(ii)$, complete $(iii)$ set of projectors of equal dimension by Eq. (\ref{qvalue}). 
We show that 
\begin{align}
\forall A\in U\mathcal{A}U^{\dagger}\,&\forall i,j\quad[\mathbf{I}_i A \mathbf{I}_j=\ket{i}\bra{j}\otimes A_{ij}]\nonumber\\
&\textrm{ with }A_{ij}A_{ij}^{\dagger}=A_{ij}^{\dagger}A_{ij}\propto\mathbb{I}_q.\label{localunitary}
\end{align}
The first line stems from the form of $I$, i.e. $\mathbf{I}_i A \mathbf{I}_j$ 
\begin{align*}
&=\sum_{pqkl} A_{pqkl} (\ket{i}\bra{i}\otimes\mathbb{I}_q)(\ket{p}\bra{q}\otimes\ket{k}\bra{l}) (\ket{j}\bra{j}\otimes\mathbb{I}_q)\\
&= \sum_{kl} A_{ijkl} (\ket{i}\bra{j}\otimes\ket{k}\bra{l}) = \ket{i}\bra{j}\otimes (\sum_{kl} A_{ijkl} \ket{k}\bra{l}). 
\end{align*}
For the second line let $M=\mathbf{I}_i A \mathbf{I}_j$. 
By Eq. (\ref{sympropto}) there is a complex number $\lambda$
such that we have both $\mathbf{I}_i=\lambda MM^{\dagger}$ 
and $\mathbf{I}_j=\lambda M^{\dagger}M$. But then
\begin{align*}
\ket{i}\bra{i}\otimes\mathbb{I}_q&=\mathbf{I}_i=\\
\lambda MM^{\dagger}&=\lambda(\ket{i}\bra{j}\otimes A_{ij})(\ket{i}\bra{j}\otimes A_{ij}^{\dagger})\\
&=\lambda\ket{i}\bra{j}\otimes A_{ij}A_{ij}^{\dagger}
\end{align*}
and hence $\lambda A_{ij}A_{ij}^{\dagger} = \mathbb{I}_q$, and symmetrically for $\lambda A_{ij}^{\dagger}A_{ij} = \mathbb{I}_q$. \vspace{1mm}\\
\textsc{Finally} consider some unitary $V=\sum_i\ket{i}\bra{i}\otimes A_{1i}$, 
where $U^{\dagger}AU$ is some matrix verifying Eq. (\ref{nondiagcase}), and rescaled so that Eq. (\ref{localunitary}) makes $A_{1j}$ it unitary. We show that 
\begin{align*}
\forall M\in VU\mathcal{A}U^{\dagger}V^{\dagger}\,&\forall i,j\quad[\mathbf{I}_i M \mathbf{I}_j=\ket{i}\bra{j}\otimes \lambda\mathbb{I}_q]\\
&\textrm{with }\lambda\textrm{ a complex number.}
\end{align*}
For a better understanding of $V$ notice that 
\begin{align*}
V&=\sum_i(\ket{i}\bra{i}\otimes A_{1i})\\
&=\sum_i (\ket{i}\bra{1}\otimes \mathbb{I}_q)(\ket{1}\bra{i}\otimes A_{1i})\\
&=\sum_i (\ket{i}\bra{1}\otimes \mathbb{I}_q) \mathbf{I}_1 A \mathbf{I}_i 
\end{align*}
Consider $B=V^{\dagger}MV$. It belongs to $U\mathcal{A}U^\dagger$ and by Eq. (\ref{localunitary}) it is of the form
$B=\sum_{ij}\ket{i}\bra{j}\otimes B_{ij}$.
Now $\mathbf{I}_i M \mathbf{I}_j$
\begin{align*}
&=\mathbf{I}_i VBV^{\dagger}\mathbf{I}_j \\
&=(\ket{i}\bra{1}\otimes \mathbb{I}_q) \mathbf{I}_1 A \mathbf{I}_i B \mathbf{I}_j A^{\dagger} \mathbf{I}_1   (\ket{1}\bra{j}\otimes \mathbb{I}_q) \\
&= (\ket{i}\bra{1}\otimes \mathbb{I}_q) \lambda \mathbf{I}_1 (\ket{1}\bra{j}\otimes \mathbb{I}_q) \\
&= \lambda (\ket{i}\bra{1}\otimes \mathbb{I}_q) (\ket{1}\bra{1}\otimes \mathbb{I}_q) (\ket{1}\bra{j}\otimes \mathbb{I}_q) \\
&= \lambda \ket{i}\bra{j}\otimes \mathbb{I}_q
\end{align*}
\hfill$\Box$

\begin{Th}[Characterizing several subsystems]~\\ \label{produit_tensoriel}
Let $\joliA$ and $\joliB$ be commuting algebras of $M_n\pa{\C}$ such that $\joliA\joliB =M_n\pa{\C}$. Then there exists a unitary matrix $U$ such that, $U\joliA U^{\dagger}$ is $M_p\pa{\C}\otimes \mathbb{I}_q$ and $U\joliB U^\dagger$ is $\mathbb{I}_p\otimes M_q\pa{\C}$, with $pq=n$.
\end{Th}
\textbf{Proof.}\\
First, let us note that $\joliC_\joliA$ includes $\C \mathbb{I}_n$. Next, the elements of $\joliC_\joliA$ commute by definition with all matrices in $\joliA$, but also with all matrices in $\joliB$, since $\joliA$ and $\joliB$ commute. Therefore, as $\joliA\joliB=M_n\pa{\C}$, $\joliC_\joliA$ is equal to $\C\mathbb{I}_n$. Thus, according to proposition~\ref{subsystem}, it is a subsystem algebra. For simplicity matters, and without loss of generality, we will assume that $\joliA$ is actually equal to $M_p\pa{C}\otimes \mathbb{I}_q$ for some $p$ and $q$ such that $pq=n$. Now for the same reasons $\joliB$ is also a subsystem algebra. Because it commutes with $\joliA$ it must act on a disjoint subsystem as $\joliA$. And since together they generate $M_n\pa{\C}$, there is no other choice but to have $\joliB$ actually equal to $\mathbb{I}_p\otimes M_q(\mathbb{C})$. \hfill$\Box$

\begin{Def}[Restriction Algebras]~\\
Consider $\mathcal{A}$ an algebra of $M_p(\mathbb{C})\otimes M_q(\mathbb{C})\otimes M_r(\mathbb{C})$.
For $A$ an element of $\mathcal{A}$, we write $A|_1$ for the matrix $\trace_{02}(A)$.
Similarly so we call $\mathcal{A}|_1$ the restriction of $\mathcal{A}$ to the middle subsystem, 
i.e. the algebra generated by the matrices of the set $\{\trace_{02}(A)\,|\,A\in\mathcal{A}\}$.
\end{Def}

\begin{Lem}[Restriction of commuting algebras]\label{commuting_restrictions}~\\
Consider $\mathcal{A}$ an algebra of $M_p(\mathbb{C})\otimes M_q(\mathbb{C}) \otimes \mathbb{I}_r$
and $\mathcal{B}$ an algebra of $\mathbb{I}_p\otimes M_q(\mathbb{C}) \otimes M_r(\mathbb{C})$.
Say $\mathcal{A}$ and $\mathcal{B}$ commute. 
Then so do $\mathcal{A}|_1$ and $\mathcal{B}|_1$.
\end{Lem}
\textbf{Proof.}\\
In the particular case where $\mathcal{A}$ and $\mathcal{B}$ have only subsystem $1$ in common we have
\begin{equation}
\forall A\in\mathcal{A}, B\in\mathcal{B}\quad pr.\trace_{02}(AB)=\trace_{02}(A)\trace_{02}(B). \label{tracemorphism}
\end{equation}
Indeed take $A=\sum_i\alpha_i.(\sigma_i\otimes\tau_i\otimes \mathbb{I})$ and $B=\sum_j\beta_j.(\mathbb{I}\otimes\mu_j\otimes\nu_j)$. We have 
\begin{align*}
pr.\trace_{02}(AB)&=\trace_{02}(\sum_{ij} pr\alpha_i\beta_j.(\sigma_i\otimes\tau_i\mu_j\otimes\nu_j))\\
&=(\sum_i p\alpha_i.\trace(\sigma_i).\tau_i)(\sum_j r\beta_j.\trace(\nu_j).\mu_j)\\
&= \trace_{02}(A)\trace_{02}(B).
\end{align*}
Now $\mathcal{A}|_1$ is generated by $\{\trace_{02}(A)\,|\,A\in\mathcal{A}\}$, and $\mathcal{B}|_1$ is generated by $\{\trace_{02}(B)\,|\,B\in\mathcal{B}\}$. Since commutation is preserved by $*$, $+$, $\alpha.$ and $\dagger$ all we need to check is that the generating elements commute. Consider $A|_1$ an element of $\mathcal{A}|_1$ and take $A$ such that $A|_1=\trace_{02}(A)$. Similarly take $B|_1$ and $B$ such that $B|_1=\trace_{02}(B)$. We have $A|_1B|_1=\trace_{02}(A)\trace_{02}(B)=pr.\trace_{02}(AB)=pr.\trace_{02}(BA)=\trace_{02}(B)\trace_{02}(A)=B|_1A|_1$. \hfill$\Box$

\begin{Lem}[Restriction of generating algebras]~\\
Consider $\mathcal{A}$ an algebra of $M_p(\mathbb{C})\otimes M_q(\mathbb{C}) \otimes \mathbb{I}_r$
and $\mathcal{B}$ an algebra of $\mathbb{I}_p\otimes M_q(\mathbb{C}) \otimes M_r(\mathbb{C})$.
Say $\mathcal{A}\mathcal{B}|_1=M_p(\mathbb{C})$. 
Hence we have that $\mathcal{A}|_1\mathcal{B}|_1=M_p(\mathbb{C})$.
\end{Lem}
\textbf{Proof.}\\
$\mathcal{A}\mathcal{B}|_1$ is generated by $\{\trace_{02}(AB)\,|\,A\in\mathcal{A}, B\in\mathcal{B} \}$. However by Eq. (\ref{tracemorphism}) this is the same as $\{\trace_{02}(A)\trace_{02}(B)\,|\,A\in\mathcal{A}, B\in\mathcal{B} \}$, which generates $\mathcal{A}|_1\mathcal{B}|_1$. \hfill$\Box$

\begin{Lem}[Duality]\label{measure}~\\
Let $\joliH_0$ and $\joliH_1$ be Hilbert spaces, with $\joliH_0$ of dimension $p$. 
Let  $A, \rho, \rho'$ denote some elements of $\joliL(\joliH_0\otimes\joliH_1)$ with $\rho, \rho'$ having 
partial traces $\rho|_0,\rho'|_0$ over $\joliH_0$.
We then have that $A\in M_p(\mathbb{C})\otimes\mathbb{I}$ is equivalent to
$$\forall\rho,\rho'\quad[\rho|_0=\rho'|_0 \Rightarrow \trace(A\rho)=\trace(A\rho')].$$
Moreover we have that $\rho|_0=\rho'|_0$ is equivalent to 
$$\forall A\in M_p(\mathbb{C})\otimes\mathbb{I} \quad[\trace(A\rho)=\trace(A\rho')].$$
\end{Lem}
\textbf{Proof.}\\
Physically the first part of the lemma says that ``a measurement is local if and only if it depends only upon the reduced density matrices''.\\
$[\Rightarrow]$. Suppose that $A=A_0\otimes\mathbb{I}_1$. In this case we have $\trace\pa{A \rho}=\trace\pa{A_0\rho|_0}$. Assuming $\rho|_0=\rho'|_0$ yields $\trace\pa{A \rho}=\trace\pa{A_0\rho|_0}=\trace\pa{A_0\rho'|_0}=\trace\pa{A \rho'}$.\\
$[\Leftarrow]$. Let's write 
$A=\sum\limits_{i,j}\ket{i}\bra{j}\otimes B_{ij}$, with $\ket{i}$ and 
$\ket{j}$ ranging over some unitary basis of $\joliH_0$. If $A$ is not 
of the form $A_0\otimes\Id_1$, then for some $i$ and $j$, $B_{ij}$ is 
not a multiple of the identity. Then there exist unit vectors $\ket{x}$ 
and $\ket{y}$ of $\joliH_1$ such that 
$\bra{x}B_{ij}\ket{x}\neq\bra{y}B_{ij}\ket{y}$. In other words, 
$\trace\pa{B_{ij}\ket{x}\bra{x}}\neq\trace\pa{B_{ij}\ket{y}\bra{y}}$. 
If we now consider $\rho=\ket{j}\bra{i}\otimes\ket{x}\bra{x}$ and 
$\rho'=\ket{j}\bra{i}\otimes\ket{y}\bra{y}$, we get what we wanted, i.e. 
$\rho|_0=\rho'|_0$ but $\trace\pa{A\rho}\neq\trace\pa{A\rho'}$.\\
Physically the second part of the lemma says that ``two reduced density matrices are the same if and only if their density matrices cannot be distinguished by a local measurement''. \\
$[\Rightarrow]$. This `$[\Rightarrow]$' is actually exactly the same as the first one, so we have already proved it.\\
$[\Leftarrow]$. Supposing $\trace\pa{A \rho}=\trace\pa{A \rho'}$ for $A=\ket{j}\bra{i}\otimes\mathbb{I}$ yields ${\rho|_0}_{ij}=\trace\pa{\ket{j}\bra{i}\rho|_0}=\trace\pa{A \rho}=\trace\pa{A \rho'}=\trace\pa{\ket{j}\bra{i}\rho'|_0}={\rho'|_0}_{ij}$. Because we can do this for all $ij$ we have $\rho|_0=\rho'|_0$.\hfill$\Box$

\section{Block structure}\label{blocks}

\noindent We will now introduce the basic definitions of one-dimensional QCA.\\

\noindent In what follows $\Sigma$ will be a fixed finite set of symbols (i.e. `the 
alphabet', describing the possible basic states each cell may 
take) and $q$ is a symbol such that $q\notin\Sigma$, which will be known as `the quiescent symbol', which represents an empty cells. We write $q\Sigma=\{q\}\cup\Sigma$ for short.\\

\begin{Def}[finite configurations]~\\ 
A \emph{(finite) configuration} $c$ of over $q\Sigma$ is a function $c:
\Z \longrightarrow q\Sigma$, with $i\longmapsto
c(i)=c_i$, such that there exists a (possibly empty)
interval $I$ verifying $i\in I\Rightarrow c_i\in q\Sigma$
and $i\notin I\Rightarrow c_i=q$. The set of 
all finite configurations over $\{q\}\cup\Sigma$ will be denoted $\mathcal{C}_f$.
\end{Def}

\noindent Whilst configurations hold the basic states of an 
entire line of cells, and hence denote the possible basic 
states of the entire QCA, the global state of a QCA may 
well turn out to be a superposition of these. The following 
definition works because $\mathcal{C}_f$ is a countably 
infinite set.\\

\begin{Def}[superpositions of configurations]~\label{superp}\\ 
Let $\mathcal{H}_{\mathcal{C}_f}$ be the Hilbert space of configurations, defined as follows. To each finite configuration $c$ is associated a unit vector $\ket{c}$, such that the family $\pa{\ket{c}}_{c\in\joliC_f}$ is an orthonormal basis of $\joliH_{\joliC_f}$. A \emph{superposition of 
configurations} is then a unit vector in $\joliH_{\joliC_f}$.
\end{Def}

\begin{Def}[Unitarity]~\label{unitarity}\\ 
A linear operator $G:\mathcal{H}_{\mathcal{C}_f}\longrightarrow\mathcal{H}_{\mathcal{C}_f}$ is \emph{unitary} if and only if $\{G\ket{c}\,|\,c\in\mathcal{C}_f\}$ is an orthonormal basis of $\mathcal{H}_{\mathcal{C}_f}.$
\end{Def}

\begin{Def}[Shift-invariance]~\label{shift-invariance}\\
Consider the shift operation which takes configuration $c=\ldots c_{i-1}c_ic_{i+1}\ldots$ to $c'=\ldots c'_{i-1}c'_ic'_{i+1}\ldots$ where for all $i$ $c'_i=c_{i+1}$. Let $\sigma:\mathcal{H}_{\mathcal{C}_f}\longrightarrow\mathcal{H}_{\mathcal{C}_f}$ be its linear extension to superpositions of configurations. A linear operator $G:\mathcal{H}_{\mathcal{C}_f}\longrightarrow\mathcal{H}_{\mathcal{C}_f}$ is said to be 
\emph{shift invariant} if and only if $G\sigma=\sigma G$.
\end{Def}

\begin{figure}
\includegraphics[scale=1.0, clip=true, trim=0cm 0cm 0cm 0cm]{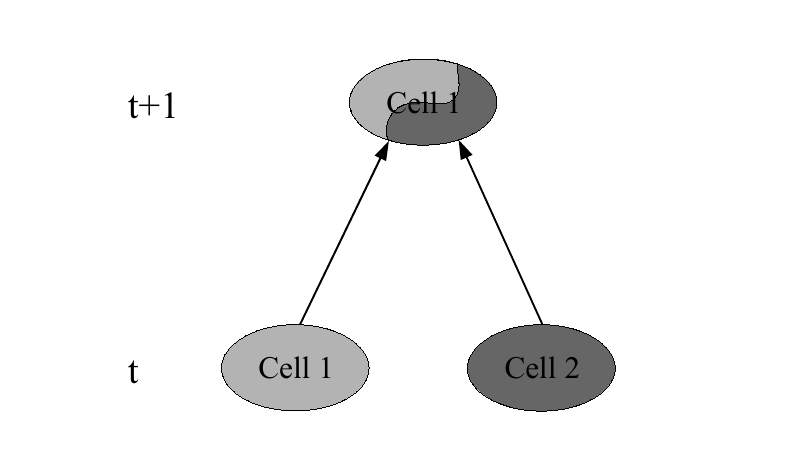}
\caption{A $\frac{1}{2}$-neighborhood CA\label{1/2}}
\end{figure}

\begin{Def}[Locality]~\label{locality}\\ 
A linear operator $G:\mathcal{H}_{\mathcal{C}_f}\longrightarrow\mathcal{H}_{\mathcal{C}_f}$ is said to be 
\emph{local} with radius $\frac{1}{2}$ if and only if for any $\rho,\rho'$ two states over $\mathcal{H}_{\mathcal{C}_f}$, and for any $i\in\Z$, we have
\begin{equation}
\rho|_{i,i+1}=\rho'|_{i,i+1}\quad \Rightarrow G\rho G^{\dagger}|_i=G\rho'G^{\dagger}|_i. \label{loceq}
\end{equation}
\end{Def}
In the classical case, the definition would be that the letter to be read in some given cell $i$ at time $t+1$ depends only the state of the cells $i$ and $i+1$ at time $t$. This seemingly restrictive definition of locality is known in the classical case as a $\frac{1}{2}$-neighborhood cellular automaton. This is because the most natural way to represent such an automaton is to shift the cells by $\frac{1}{2}$ at each step, so that the state of a cell depends on the state of the two cells under it, as shown in figure~\ref{1/2}. This definition of locality is actually not so restrictive, since by grouping cells into `supercells' one can construct a $\frac{1}{2}$-neighborhood CA simulating the first one. The same thing can easily be done for QCA, so that this definition of locality is essentially done without loss of generality. Transposed to a quantum setting, we get the above definition: to know the state of cell number $i$, we only need to know the state of cells $i$ and $i+1$ before the evolution.\\
\noindent We are now set to give the formal definition of one-dimensional quantum cellular automata.
\begin{Def}[QCA]~\label{lca}\\ A one-dimensional quantum cellular 
automaton (QCA) is an operator $G:\mathcal{H}_{\mathcal{C}_f}\longrightarrow\mathcal{H}_{\mathcal{C}_f}$
which is unitary, shift-invariant and local.
\end{Def}
The next theorem provides us with another characterization of locality, more helpful in the proofs. But more importantly it entails structural reversibility, i.e. the fact that the inverse function of a QCA is also a QCA. Actually this theorem works for $n$-dimensional QCA as well as one.
\begin{Th}[Structural reversibility]~\\ \label{localization}
Let $G$ be a unitary operator of $\joliH_{C_f}$ and $\joliN$ a finite 
subset of $\Z$. The two properties are equivalent:
\begin{itemize}
\item[(i)] For every states $\rho$ and $\rho'$ over the finite 
configurations, if $\rho|_\joliN=\rho'|_\joliN$ then $\pa{G\rho 
G^\dagger}|_0=\pa{G\rho' G^\dagger}|_0$.\\
\item[(ii)] For every operator $A$ localized on cell $0$, then $G^\dagger 
AG$ is localized on the cells in $\joliN$.\\
\item[(iii)] For every states $\rho$ and $\rho'$ over the finite 
configurations, if $\rho|_{-\joliN}=\rho'|_{-\joliN}$ then $\pa{G^\dagger\rho 
G}|_0=\pa{G^\dagger\rho' G}|_0$.\\
\item[(iv)] For every operator $A$ localized on cell $0$, then $G 
AG^\dagger$ is localized on the cells in $-\joliN$.
\end{itemize}
When $G$ satisfies these properties, we say that $G$ is local at $0$ 
with neighbourhood $\joliN$.
\end{Th}
\textbf{Proof.}\\
$[(i)\Rightarrow(ii)]$. Suppose (i) and let $A$ be an operator 
acting on cell $0$. For every states $\rho$ and $\rho'$ such that 
$\rho|_\joliN=\rho'|_\joliN$, we have $\trace\pa{AG\rho 
G^\dagger}=\trace\pa{AG\rho' G^\dagger}$, using lemma \ref{measure} and our hypothesis that $\pa{G\rho G^\dagger}|_0=\pa{G\rho' G^\dagger}|_0$. We thus 
get $\trace\pa{G^\dagger AG\rho}=\trace\pa{G^\dagger AG\rho'}$. Since 
this is true of every $\rho$ and $\rho'$ such that 
$\rho|_\joliN=\rho'|_\joliN$, this means, again according to 
lemma \ref{measure}, that $G^\dagger AG$ is localized on the cells in $\joliN$.\\
$[(ii)\Rightarrow (i)]$. Suppose (ii) and $\rho|_\joliN=\rho'|_\joliN$. 
Then, for every operator $B$ localized on the cells in $\joliN$, 
lemma \ref{measure} gives $\trace\pa{B\rho}=\trace\pa{B\rho'}$, 
so for every operator $A$ localized on cell 0, we get:
\begin{align*}
\trace\pa{AG\rho G^\dagger}&=\trace\pa{G^\dagger 
AG\rho}\\
&=\trace\pa{G^\dagger AG\rho'}\\
&=\trace\pa{AG\rho'G^\dagger}
\end{align*}
Again by lemma \ref{measure}, this means $\pa{G\rho G^\dagger}|_0=\pa{G\rho' G^\dagger}|_0$.\\
$[(ii)\Rightarrow (iv)]$. Suppose $(ii)$ and let $A$ be an operator acting on cell $0$. Consider some operator $M$ acting on a cell $i$ which does not belong to $-\joliN$. According to our hypothesis we know that $G^{\dagger}MG$ does not act upon cell $0$, and hence it commutes with $A$. But $AB\mapsto GAG^{\dagger}GBG^{\dagger}=GABG^{\dagger}$ is a morphism, hence $GG^{\dagger}MGG^{\dagger}=M$ also commutes with $GAG^{\dagger}$. Because $M$ can be chosen amongst to full matrix algebra $M_d(\mathbb{C})$ of cell $i$, this entails that $GAG^{\dagger}$ must be the identity upon this cell. The same can be said of any cell outside $-\joliN$.\\
$[(iv)\Rightarrow (ii)]$, $[(iii)\Rightarrow (iv)]$, $[(iii)\Leftarrow (iv)]$ are symmetrical to 
$[(ii)\Rightarrow (iv)]$, $[(i)\Rightarrow (ii)]$, $[(ii)\Leftarrow (i)]$ just by interchanging the roles of $G$ and $G^{\dagger}$.
\hfill $\Box$

\begin{figure}
\includegraphics[scale=1.0, clip=true, trim=0cm 0cm 0cm 0cm]{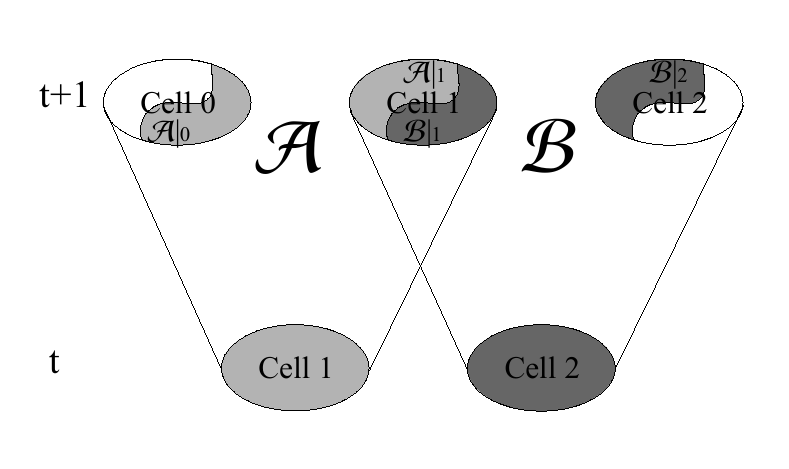}
\caption{Definitions of the algebras for the proof of the structure theorem.\label{ABalgebras} }
\end{figure}

\noindent Now this is done we proceed to prove the structure theorem for QCA over finite, unbounded configurations. This is a simplification of \cite{Werner}. The basic idea of the proof is that in a cell at time $t$ we can separate what information will be sent to the left at time $t+1$ and which information will be sent to the right at time $t+1$. But first of all we shall need two lemmas. These are better understood by referring to Figure \ref{ABalgebras}.
\begin{Lem}\label{separation}
Let $\mathcal{A}$ be the image of the algebra of the cell $1$ under the global evolution $G$. It is localized upon cells $0$ and $1$, and we call $\mathcal{A}|_1$ the restriction of $\mathcal{A}$ to cell $1$.\\
Let $\mathcal{B}$ be the image of the algebra of the cell $2$ under the global evolution $G$. It is localized upon cells $1$ and $2$, and we call $\mathcal{B}|_1$ the restriction of $\mathcal{B}$ to cell $1$.\\
There exists a unitary $U$ acting upon cell $1$ such that $U\mathcal{A}|_0U^{\dagger}$ is of the form $M_p(\mathbb{C})\otimes \mathbb{I}_q$ and $U\mathcal{B}|_1U^{\dagger}$ is of the form $\mathbb{I}_p\otimes M_q(\mathbb{C})$, with $pq=d$.
\end{Lem}
\textbf{Proof.}\\ 
$\mathcal{A}$ and $\mathcal{B}$ are indeed localized as stated due to the locality of $G$ and a straightforward application of lemma \ref{localization} with $\joliN=\{0,1\}$, which we can apply at position $1$ and $2$ by shift-invariance.\\
$\mathcal{A}$ and $\mathcal{B}$ commute because they are the image of two commuting algebras, those of Cell $1$ and $2$, via a morphism $AB\mapsto GAG^{\dagger}GBG=GABG^{\dagger}$.\\
Moreover by lemma \ref{localization} the antecedents of the operators localized in cell $1$ are all localized in cells $1$ and $2$. Plus they all have an antecedent because $G$ is surjective. Hence $\mathcal{A}\mathcal{B}|_1$ is the entire cell algebra of cell $1$, i.e. $M_d(\mathbb{C})$.\\
So now we can apply Theorem \ref{produit_tensoriel} and the result follows. \hfill$\Box$
 
\begin{Lem}\label{inclusion}~\\
Let $\mathcal{B}$ be the image of the algebra of the cell $2$ under the global evolution $G$. It is localized upon cells $1$ and $2$, and we call $\mathcal{B}|_1$ the restriction of $\mathcal{B}$ to cell $1$ and $\mathcal{B}|_2$ the restriction of $\mathcal{B}$ to cell $2$.\\
We have that  $\mathcal{B}=\mathcal{B}|_1\otimes\mathcal{B}|_2$.
\end{Lem}
\textbf{Proof.}\\
We know that $\mathcal{B}$ is isometric to $M_d(\mathbb{C})$ and we know that 
$\mathcal{B}|_1\otimes\mathcal{B}|_2\subset\mathcal{B}$.
But then by the previous lemma applied upon cell $1$ we also know that $\mathcal{B}|_1$ is isometric to $M_q(\mathbb{C})$ and if we apply it to cell $2$ then we have that $\mathcal{B}|_2$ is isometric to $M_p(\mathbb{C})$.
Hence the inclusion is an equality.
\hfill$\Box$

\begin{figure}
\includegraphics[scale=1.0, clip=true, trim=0cm 0cm 0cm 0cm]{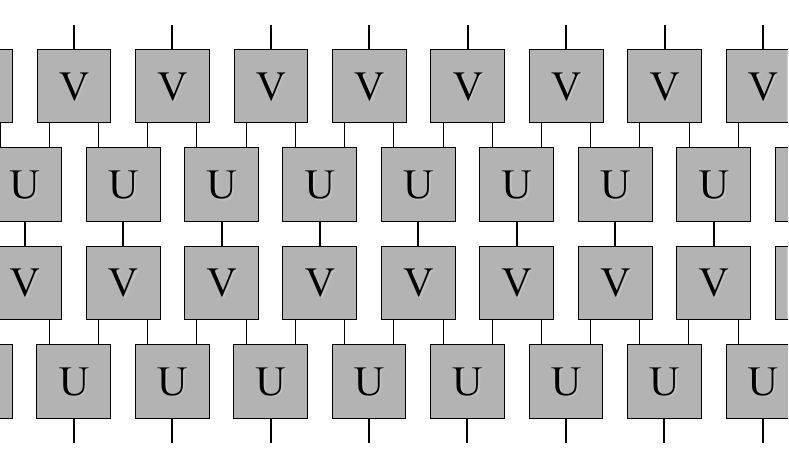}
\caption{QCA with two-layered block representation $(U,V)$.\label{structure} Each line represents a cell, which is a quantum system. Each square represents a unitary $U/V$ which gets applied upon the quantum systems. Time flows upwards.}
\end{figure}

\begin{figure}
\includegraphics[scale=1.0, clip=true, trim=0cm 0cm 0cm 0cm]{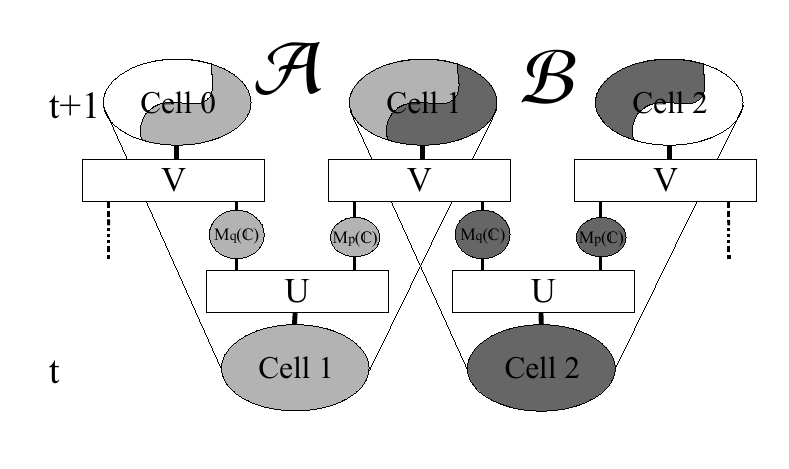}
\caption{Zooming into the two-layered block representation.\label{werner} The unitary interactions $U$ and $V$ are alternated repeatedly as shown.}
\end{figure}

\begin{Th}[Structure theorem]~\\ \label{coeur}
Any QCA $G$ is of the form described by Figures \ref{structure} and 
\ref{werner}.
\end{Th}
\textbf{Proof.}\\
Let $\joliA$ and $\joliB$ be respectively the images of 
the algebra of the cells $1$ and $2$ under the global evolution $G$. By 
virtue of lemma \ref{separation} we know that $\joliA$ and $\joliB$ are 
respectively isometric to $M_p\pa{\C}\otimes \Id_q$ and $\Id_p\otimes 
M_q\pa{\C}$; let $V^{\dagger}$ be the unitary transformation over $\C^d$ 
which 
accomplishes this separation. From lemma~\ref{inclusion}, we know that 
$(V^{\dagger}\otimes 
V^{\dagger})G$ maps the algebra of one cell into $\Id_p\otimes 
M_q(\C)\otimes 
M_p(\C)\otimes \Id_q$, so we can choose a unitary operator $U$ over 
$\C^d$ 
which realizes this mapping by conjugation. By shift-invariance, the 
same $V$ and $U$ will do for every position in the line.
Therefore $G=\pa{\bigotimes V}\pa{\bigotimes U}$ as in Fig. 
\ref{werner}. 
The rest of the proof serves only to give a formal meaning to these 
infinite
tensor products as unitary operators over $\joliH_{\joliC_f}$.\\
Let us consider $\ket{q}\bra{q}\in M_d\pa{\C}$. Its image by 
$V^{\dagger}$, i.e. 
$V\ket{q}\bra{q}V^\dagger$, is some one-dimensional projector in 
$M_p(\C)\otimes M_q(\C)$. Now consider the state corresponding to the 
quiescent 
state on every cells. It is invariant by $G$, so this 
$V\ket{q}\bra{q}V^\dagger$
has to be separable in $M_p\pa{\C}\otimes M_q\pa{\C}$, because after 
applying independent $U$
transformations on each side we get the everywhere quiescent state, 
which is unentangled. This means that $V\ket{q}\bra{q}V^\dagger$ can be 
written as $\ket{q_1}\bra{q_1}\otimes\ket{q_2}\bra{q_2}$, where 
$\ket{q_1}$ and $\ket{q_2}$ are respectively unit vectors of $\C^p$ and 
$\C^q$. So we can assume that $V$ maps $\ket{q_1}\ket{q_2}$ to 
$\ket{q}$.
Moreover we know 
$U^{\dagger}\pa{\ket{q_2}\bra{q_2}\otimes\ket{q_1}\bra{q_1}}U$ 
must be equal to $\ket{q}\bra{q}$, so we can assume that $U$ maps 
$\ket{q}$ to $\ket{q_2}\ket{q_1}$.\\
We can now give a meaning to the infinite product of unitary operators. 
For each $n$ we consider the operator $\pa{\bigotimes_{[-n,n]} U}$  
where the $U$'s are only applied on the portion $[-n,n]$ of the line. 
The action of $\pa{\bigotimes U}$ is simply the limit of its images 
by $\pa{\bigotimes_{[-n,n]} U}$, when $n$ goes to infinity. 
That $U$ maps $\ket{q}$ to $\ket{q_2}\ket{q_1}$ insures that this limit 
does exist. Indeed, for every finite configuration $c$, the sequence 
$\pa{\bigotimes_{[-n,n]} U}\ket{c}$ will be ultimately constant, 
due to the quiescent boundaries.\hfill$\Box$\\
Note that this structure could be further simplified if we were to allow ancillary cells \cite{Arrighi1}.

\section{Quantizations and consequences}\label{discussion}

\noindent The structure theorem for QCA departs in several important ways from the classical situation, giving rise to a number of apparent paradoxes. We begin this section by discussing some of these concerns in turns. Each of them is introduced via an example, which we then use to derive further consequences or draw the limits of the structure theorem.\vspace{1mm}

\noindent \emph{Bijective CA and superluminal signalling.}\\
First of all, it is a well-known fact that not all bijective CA are structurally reversible. 
The XOR CA is a standard example of that.
\begin{Def}[XOR CA]~\\
Let $\mathcal{C}_f$ be the set of finite configurations over the alphabet $q\Sigma =\{q,0,1\}$.
For all $x,y$ in $q\Sigma$ Let $\delta(qx)=x,\,\delta(xq)=q$, and $\delta(xy)=x\oplus y$ otherwise.
We call $F : \mathcal{C}_f \longrightarrow \mathcal{C}_f$ the function mapping $c=\ldots c_{i-1} c_i c_{i+1} \ldots$ to $c'=\ldots \delta(c_{i-1} c_i)\delta(c_{i} c_{i+1})\ldots$.
\end{Def}
The XOR CA is clearly shift-invariant, and local in the sense that the state of a cell at $t+1$ only depends from its state and that of its right neighbour at $t$. It is also bijective. Indeed for any $c'=\ldots qqc'_kc'_{k+1}\ldots$ with $c'_k$ the first non quiescent cell, we have $c_k=q$, $c_{k+1}=c'_k$, and thereon for $l\geq k+1$ we have either $c_{l+1}=c_l\oplus c'_l$ if $c'_l\neq q$, or once again $c_{l+1}=q$ otherwise, etc. In other words the antecedent always exists (surjectivity) and is uniquely derived (injectivity) from left till right. But the XOR CA is not structurally reversible. Indeed for some $c'=\ldots 000000000 \ldots$ we cannot know whether the antecedent of this large zone of zeroes is another large zone of zeroes or a large zone of ones -- unless we deduce this from the left border as was previously described\ldots but the left border may lie arbitrary far.\\
So classically there are bijective CA whose inverse is not a CA, and thus who do not admit any $n$-layered block representation at all. Yet surely, just by defining $F$ over $\mathcal{H}_{\mathcal{C}_f}$ by linear extension (e.g. $F(\alpha.\ket{\ldots 01 \ldots}+\beta.\ket{\ldots 11 \ldots})=\alpha.F\ket{\ldots 01 \ldots}+\beta.F\ket{\ldots 11 \ldots})$ we ought to have a QCA, together with its block representation, hence the apparent paradox.\\
In order to lift this concern let us look at the properties of this quantized 
$F : \mathcal{H}_{\mathcal{C}_f} \longrightarrow \mathcal{H}_{\mathcal{C}_f}$.
It is indeed unitary as a linear extension of a bijective function, and it is shift-invariant for the same reason. Yet counter-intuitively it is non-local. Indeed consider configurations $c_\pm=1/\sqrt{2}.\ket{\ldots qq}(\ket{00\ldots 00}\pm\ket{11\ldots 11})\ket{qq\ldots}$. We have $F c_\pm = \ket{\ldots qq00\ldots 0}\ket{\pm}\ket{qq\ldots}$, where we have used the usual notation $\ket{\pm}=1/\sqrt{2}.(\ket{0}\pm\ket{1})$. Let $i$ be the position of this last non quiescent cell. Clearly $(F c_{\pm})|_i=\ket{\pm}\bra{\pm}$ is not just a function of $c|_{i,i+1}=(\ket{0q}\bra{0q}+\ket{1q}\bra{1q})/2$, but instead depends upon this global $\pm$ phase. Another way to put it is that the quantized XOR may be used to transmit information faster than light. Say the first non quiescent cell is with Alice in Paris and the last non quiescent cell is with Bob in New York. Just by applying a phase gate $Z$ upon her cell Alice can change $c_{+}$ into $c_{-}$ at time $t$, leading to a perfectly measurable change from $\ket{+}$ to $\ket{-}$ for Bob. Again another way to say it is that operators localized upon cell $1$ are not taken to operators localized upon cells $0$ and $1$, as was the case for QCA. For instance take $\mathbb{I}\otimes Z\otimes \mathbb{I}$ localized upon cell $1$. This is taken to $F(\mathbb{I}\otimes Z\otimes \mathbb{I})F^{\dagger}$. But this operation is not localized upon cells $0$ and $1$, as it takes $\ket{\ldots qq00\ldots 0}\ket{+}\ket{qq\ldots}$ to $\ket{\ldots qq00\ldots 0}\ket{-}\ket{qq\ldots}$, whatever the position $i$ of the varying $\ket{\pm}$.\\
Now let us take a step back. If a CA is not structurally reversible, there is no chance that its QCA will be. Moreover according the current state of modern physics, quantum mechanics is the theory for describing all closed systems. Therefore we reach the following proposition.
\begin{Pro}[Class $B$ is not locally quantizable]
The class of bijective but not structurally reversible CA upon finite configurations is known to coincide with the class of surjective but non injective CA upon infinite configurations, or again the class of bijective CA upon finite configurations but not upon infinite configuration. Call it $B$.\\ 
The quantization of a class $B$ automata is not local. It cannot be implemented by a series of finite closed quantum systems.

\end{Pro}
As far as CA are concerned this result removes much of the motivation of several papers which focus upon class $B$. As regards QCA the structure theorem removes much of the motivation of the papers \cite{Durr1, Durr2, Arrighi1}, which contain unitary decision procedures for possibly non-structurally reversible QCA.\\

\begin{figure}
\includegraphics[scale=1.0, clip=true, trim=0cm 0cm 0cm 0cm]{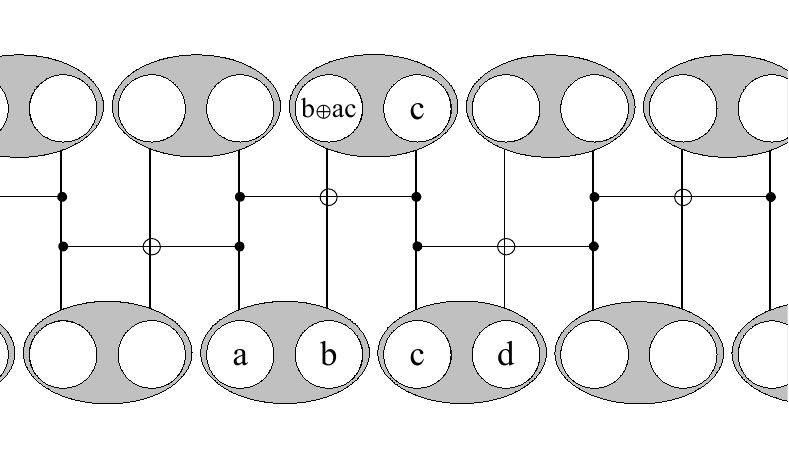}
\caption{The Toffoli CA.\label{toffoli}}
\end{figure}

\noindent \emph{Faster quantum signalling.}\\
Second, it is a well-known fact that there exists some radius $1/2$, structurally reversible CA, whose inverse is also of radius $1/2$, and yet which do not admit a two-layered block representation unless the cells are grouped into supercells. The Toffoli CA is a good example of that.
\begin{Def}[Toffoli CA]
Let $\mathcal{C}_f$ be the set of finite configurations over the alphabet $\{00,01,01,11\}$, with $00$ now taken as the quiescent symbol. For all $ab$ and $cd$ taken in the alphabet let $\delta(abcd)= (b\oplus a.c)c$.
We call $F : \mathcal{C}_f \longrightarrow \mathcal{C}_f$ the function mapping $c=\ldots c_{i-1} c_i c_{i+1} \ldots$ to $c'=\ldots \delta(c_{i-1} c_i)\delta(c_{i} c_{i+1})\ldots$. This is best described by Figure \ref{toffoli}.
\end{Def}
The Toffoli CA is clearly shift-invariant, and of radius $1/2$. Let us check that its inverse is also of radius $1/2$. For instance say we seek to retrieve (c,d). $c$ is easy of course. By shift-invariance retrieving $d$ is like retrieving $b$. But since we have $a$ and $d$ in cleartext we can easily substract $a.d$ from 
$b\oplus a.d$. 
Now why does it not have a two-layered block representation without cell grouping? Remember the toffoli gate is the controlled-controlled-NOT gate. Here $b$ is NOTed depending upon $a$ and $c$, which pass through unchanged, same for $d$ with the left and right neighbouring subcells, etc. So actually the Toffoli CA is just two layers of the toffoli gate, as we have shown in Figure \ref{toffoli}. 
But we know that the toffoli gate cannot be obtained from two bit gates in classical reversible electronics, hence there cannot be a two-layered block representation without cell grouping. \\
So classically there exists some structurally reversible CA, of radius $1/2$, whose inverse is also of radius $1/2$, but do not admit a two-layered block representation without cell grouping. Yet surely, just by defining $F$ over $\mathcal{H}_{\mathcal{C}_f}$ by linear extension we ought to have a QCA, together with its block representation, and that construction does not need any cell grouping, hence again the apparent paradox.\\
Again in order to lift this concern let us look at the properties of this quantized 
$F : \mathcal{H}_{\mathcal{C}_f} \longrightarrow \mathcal{H}_{\mathcal{C}_f}$.
It is indeed unitary and shift-invariant of course. This time it is also local, but counter-intuitively it turns out not to be of radius $1/2$. Indeed from the formulation in terms of Toffoli gates as in Figure \ref{toffoli} one can show that the radius is $3/2$ in a quantum mechanical setting. For instance one can check that putting $\ket{+}$ in the $a$-subcell, $\ket{-}$ in the $b$-subcell, and either $\ket{0}$ or $\ket{1}$ in the $c$-subcell of Fig. \ref{toffoli} at time $t$ will yield either $\ket{+}$ or $\ket{-}$ in the $a$-subcell at time $t+1$.\\
Once more let us take a step back. The Toffoli CA is yet another case where exploiting quantum superpositions of configurations enables us to have information flowing faster than in the classical setting, just like for the XOR CA. But unlike the XOR CA, the speed of information remains bounded in the Toffoli CA, and so up to cell grouping it can still be considered a QCA. Therefore we reach the following proposition.
\begin{Pro}[Quantum information flows faster]
Let $F : \mathcal{C}_f \longrightarrow \mathcal{C}_f$ be a CA and $F : \mathcal{H}_{\mathcal{C}_f} \longrightarrow \mathcal{H}_{\mathcal{C}_f}$ the corresponding QCA, as obtained by linear extension of $F$. Information may flow faster in the the quantized version of $F$.
\end{Pro}
This result is certainly intriguing, and one may wonder whether it might contain the seed of a novel development quantum information theory, as opposed to its classical counterpart.\vspace{1mm}

\begin{figure}
\includegraphics[scale=1.0, clip=true, trim=0cm 0cm 0cm 0cm]{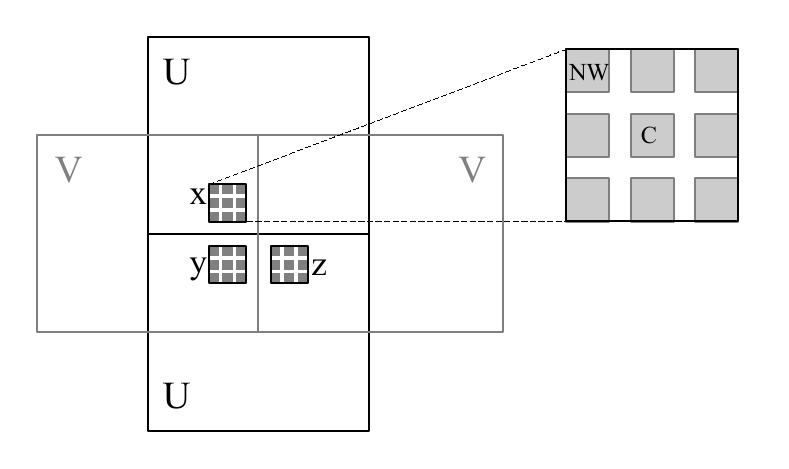}
\caption{The Kari CA and the choice of $x,y,z$ cells.\label{kari0}}
\end{figure}

\noindent \emph{No-go for $n$-dimensions.}\\
Finally, it is again well-known that in two-dimensions there exists some structurally reversible CA which do not admit a two-layered block representation, even after a cell-grouping. The standard example is that of Kari \cite{Kari}:
\begin{Def}[Kari CA]
Let $\mathcal{C}_f$ be the set of finite configurations over the alphabet $\{0,1\}^9$, with $0^{9}$ is now taken as the quiescent symbol. So each cell is made of $8$ bits, one for each cardinal direction (North, North-East\ldots) plus one bit in the center, as in Figure \ref{kari0}. At each time step, the North bit of a cell undergoes a $NOT$ only if the cell lying North has center bit equal to $1$, the North-East bit of a cell undergoes a $NOT$ only if the cell lying North-East has center bit equal to $1$, and so on. Call $F$ this CA.
\end{Def}
\begin{figure}
\includegraphics[scale=1.0, clip=true, trim=0cm 0cm 0cm 0cm]{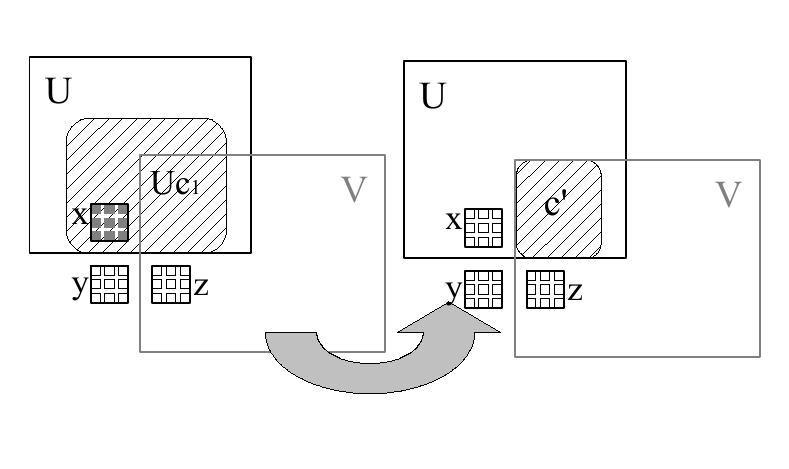}
\caption{$Uc_1$ and $c'$.\label{kari1}}
\end{figure}
\begin{figure}
\includegraphics[scale=1.0, clip=true, trim=0cm 0cm 0cm 0cm]{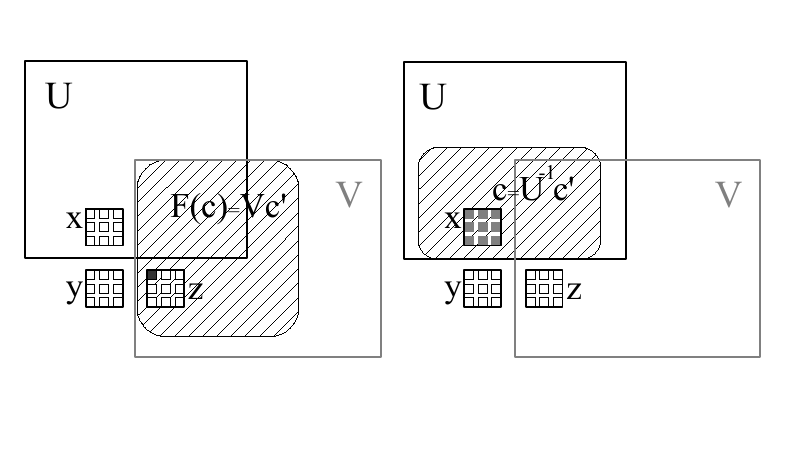}
\caption{$F(c)$ and $c$: a contradiction.\label{kari2}}
\end{figure}
This is clearly shift-invariant, local and structurally reversible. Informally the proof that the Kari CA does not admit a a two-layered block representation, even if we group cells into supercells, runs as follows \cite{Kari}:\\
- Suppose $F$ admits a decomposition into $U$ and $V$ blocks;\\
- Consider cells $x,y,z$ such that they are all neighbours; $x,y$ are in the same $V$-block but not $z$; $x$ and $y,z$ are not in the same $U$-block, as in Figure \ref{kari0};\\
- Consider $c_1$ the configuration with $1$ as the center bit of $x$ and zero everywhere. Run the $U$-blocks, the hatched zone left of Figure \ref{kari1} represents those cells which may not be all zeros;\\
- Consider the configuration obtained from the previous $Uc_1$ by putting to zero anything which lies not the $V$-block of $z$, as in Figure \ref{kari1}. Call it $c'$, and let $c$ be defined as the antecedent of $c$ under a run of the $U$-blocks;\\
- On the one hand we know that $F(c)$ is obtained from a run of the $V$-blocks from $c'$. In the left $V$-block there are just zeroes so $F(c)$ has only zeroes, in particular the North bit of the $y$ cell is $0$. But the right $V$-block is exactly the same as that of $Uc_1$, and so we know that the North-West bit of the $y$ cell of $F(c)$ is $1$, as in the left of Figure \ref{kari2};\\
- On the other hand since $c'$ has zeroes everywhere but in the above $U$-block, the same is true of $c$, as shown right of the Figure \ref{kari2}. A consequence of this is that the North bit of the $y$ cell must be equal to the North-West bit of the $z$ cells, as they are both obtained from the same function ofs the center bit of the $x$ cell;\\
- Hence the contradiction.\\
Now by defining $F$ over $\mathcal{H}_{\mathcal{C}_f}$ by linear extension we have a QCA, and the proof applies in a very similar fashion, the contradiction being that the state of the North qubit of the $y$ cell must be equal to the state of the North-West bit of the $z$ cells, whether these are mixed states or not. Hence we have a counterexample to the higher-dimensional case of the Theorem in \cite{Werner}. We reach the following proposition.
\begin{Pro}[No-go for $n$-dimensions]
There exists some $2$-dimensional QCA which do not admit a two-layered block representation.
\end{Pro}
Understanding the structure of the $n$-dimensional QCA is clearly the main challenge that remains ahead of us.

\section*{Acknowledgements}
We would like to thank Jacques Mazoyer, Torsten Franz, Holger Vogts, Jarkko Kari, J\'er\^ome Durand-Lose, Renan Fargetton, Philippe Jorrand for a number of helpful conversations.


\begin{thebibliography}{99}
\bibitem{Arrighi1} P. Arrighi, \emph{Intrinsically universal one-dimensional quantum cellular automata}, MFCS'07, (2007).
\bibitem{Arrighi2} P. Arrighi, R. Fargetton, \emph{Intrinsically universal one-dimensional quantum cellular automata}, DCM'07, (2007).
\bibitem{Bratteli} O. Bratteli, D. Robinson, \emph{Operators algebras and quantum statistical mechanics 1}, Springer, (1987).
\bibitem{Durr1} C. D\"urr, H. L\^eThanh, M. Santha, \emph{A decision procedure for well formed quantum cellular
automata}, Random Structures and Algorithms, $\mathbf{11}$,
381--394, (1997).
\bibitem{Durr2} C. D\"urr, M. Santha, \emph{A decision procedure for unitary quantum linear cellular
automata}, SIAM J. of Computing, $\mathbf{31}(4)$, 1076--1089,
(2002).
\bibitem{Feynman} R. P. Feynman, \emph{Quantum mechanical computers}, Found. Phys. 
$\mathbf{16}$, 507-531, (1986).
\bibitem{Gijswijt} D. Gijswijt, \emph{Matrix algebras and semidefinite programming techniques for codes}, Ph.D. thesis, University of Amsterdam, 2005.
\bibitem{Kari} J. Kari, \emph{On the circuit depth of structurally reversible cellular automata}, Fudamenta Informaticae, $\mathbf{34}$, 1-15, (1999).
\bibitem{Werner} B. Schumacher, R. F. Werner, \emph{Reversible quantum cellular automata}, 
arXiv:quant-ph/0405174. 
\bibitem{Shepherd} D. J. Shepherd, T. Franz, R. F. Werner, \emph{Universally programmable quantum cellular automata}, Phys. Rev. Lett., $\mathbf{97}$, 020502 (2006).
\bibitem{Watrous} J. Watrous, \emph{On one-dimensional quantum cellular 
automata}, Complex Systems $\mathbf{5}(1)$, 19--30, (1991).
\end{thebibliography}
\end{document}